\documentclass[codesnippet, shortnames, nojss]{jss}
\usepackage{amsmath, amssymb}

\title{The \pkg{Langevin} Approach:\\ 
       An \proglang{R} Package for Modeling Markov Processes}

\author{Philip Rinn\\Universit\"at Oldenburg \And 
        Pedro G. Lind\\Universit\"at Oldenburg \And 
        Matthias W\"achter\\Universit\"at Oldenburg \And 
        Joachim Peinke\\Universit\"at Oldenburg}

\Plainauthor{Philip Rinn, Pedro Lind, Matthias W\"achter, Joachim Peinke}
\Plaintitle{The Langevin Approach: An R Package for Modeling Markov Processes}

\Abstract{We describe an \proglang{R} package developed by the research group \textit{Turbulence, Wind energy and Stochastics} (TWiSt) at the Carl von Ossietzky University of Oldenburg, which extracts the (stochastic) evolution equation underlying a set of data or measurements. The method can be directly applied to data sets with one or two stochastic variables. Examples for the one-dimensional and two-dimensional cases are provided. This framework is valid under a small set of conditions which are explicitly presented and which imply simple preliminary test procedures to the data. For Markovian processes involving Gaussian white noise, a stochastic differential equation is derived straightforwardly from the time series and captures the full dynamical properties of the underlying process. Still, even in the case such conditions are not fulfilled, there are alternative versions of this method which we discuss briefly and provide the user with the necessary bibliography.}

\Plainkeywords{R, stochastic processes, data analysis}

\Address{Philip Rinn,\\
  Institute of Physics and\\
  ForWind -- Center for Wind Energy research\\
  C.v.O. Universit\"at Oldenburg\\
  26111 Oldenburg, Germany\\
  E-mail: \email{philip.rinn@uni-oldenburg.de}\\
  URL: \url{http://www.uni-oldenburg.de/twist}}

\begin{document}

\section{Introduction}

When dealing with stochastic series of data measurements, standard statistical tools, such as mean and centered moments, are able to catch the essential features of the distribution of observed values. Last end, sufficient high-order moments will retrieve a good approximation of the probability density function (PDF) associated with the stochastic process. However, PDFs are not able to fully characterize the dynamics underlying the process. A typical example is the Gaussian distribution: if the stochastic variable assumes values according to a Gaussian distribution, the dynamics producing such distribution of values can be as simple as an Ornstein-Uhlenbeck process \citep{Risken1996} but it may also be the result of a much more complicated dynamics, as we exemplify below. Thus, while knowing the distribution of observed values is important as a first approach to the data, uncovering the complete dynamics of the process provides a much deeper insight into the system, which cannot be accessed through standard statistical tools.

Starting from a stochastic differential equation, a process can be statistically reconstructed through simple stochastic integration. The inverse problem however is much more complicated: would a set of measurements be enough for a bottom-up approach to infer the underlying dynamics of the process? The short answer is yes, there are cases where this is possible. In this paper we present the long answer implemented as a package for \proglang{R} \citep[see][]{RCT2015}, which can be easily used, composing a method which we call the Langevin Approach. This approach was introduced by Peinke and Friedrich in the late 1990s \citep{Friedrich1997,Siegert1998} and further developed in the last decades. For a review see \citet{Friedrich2011}.

We start in Section~\ref{sec:maths} by describing the mathematical background on which our method is based. Here we also discuss the necessary conditions for the method to be applicable. The \proglang{R} functions implementing our approach are presented in Section~\ref{sec:implementation} together with two examples for one and for two stochastic variables. Additional remarks on our method are discussed in Section~\ref{sec:remarks}, providing the user with supplementary issues and suggestions for more general situations, in particular those where the conditions supporting our mathematical background do no longer hold. A selected bibliography is also provided. Finally, Section~\ref{sec:conclusions} concludes the paper.

\section{Stochastic equations: from data to models and back}
\label{sec:maths}

\subsection{The Langevin model}

A wide range of dynamical systems can be described by a stochastic differential equation, the (non-linear) Langevin equation \citep[cf.][]{Risken1996,Hanggi1982,Kampen2007}. 

Consider a general stochastic trajectory $X(t)$ in time $t$. The time derivative of the system's trajectory $\frac{dX}{dt}$ can be expressed as the sum of two complementary contributions: one being purely deterministic and another one being stochastic, governed by a stochastic ``force'' $\Gamma(t)$, defined as a $\delta$-correlated Gaussian white noise, i.e., $\langle \Gamma(t) \rangle = 0$ and $\langle \Gamma(t) \Gamma(t^{\prime}) \rangle = 2\delta(t - t^{\prime})$.
While the deterministic term is defined by a function, $D^{(1)}(X)$ the stochastic contribution is weighted by another function, $D^{(2)}(X)$, yielding the evolution equation of $X$

\begin{equation}
\frac{dX}{dt} = D^{(1)}(X) + \sqrt{D^{(2)}(X)} \, \Gamma(t) \, ,
\label{Langevin}
\end{equation} 

where the square root is taken for consistency, as will be clear below. We assume stationary time series here, so $D^{(1)}$ and $D^{(2)}$ are not time dependent but we show briefly how non-stationary time series can be treated in Section~\ref{sec:remarks}.

The Langevin equation should be interpreted as follows: for every time $t$ where the system meets an arbitrary but fixed point $X$ in phase space, $X(t + \tau)$ with small $\tau$ is defined by the deterministic function $D^{(1)}(X)$ and the stochastic function $\sqrt{D^{(2)}(X)}\Gamma(t)$, through trivial (Euler) stochastic integration \citep{Friedrich2011}:

\begin{equation}
X(t + \tau) = X(t) + D^{(1)}(X)\tau + \sqrt{D^{(2)}(X)\tau} \, \eta(t) \, ,
\label{LangevinDisc}
\end{equation}

where $\eta(t)$ is a normally distributed random variable. Here we use the It\^o picture of stochastic integrals, for further details see \citet{Gardiner2004}.

%%%%%%%%%%%%%%%%%%%%%%%%%%%%%%%%%%%%%%%%%%%%%%%%%%%%%%%%%
\begin{figure}
\centering
\includegraphics[width=\textwidth]{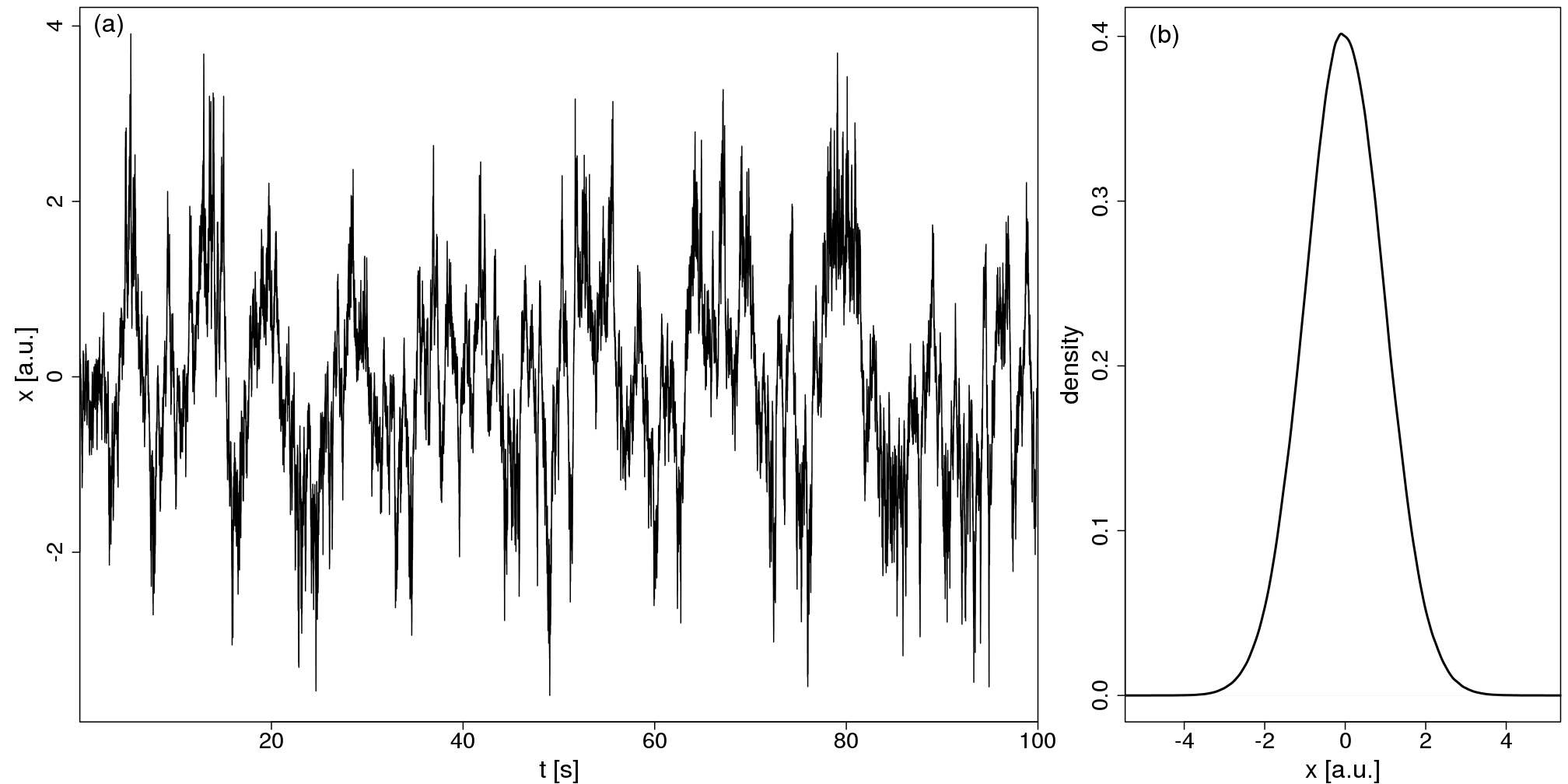}
\caption{{\bf (a)} Sketch of a stochastic process in time governed by a cubic drift and quadratic diffusion contributions and {\bf (b)} its corresponding probability density function (PDF). Though the series shows a bistable dynamics (cubic drift) the PDF follows a Gaussian function, equivalent to an Ornstein-Uhlenbeck process (see text).}
\label{fig01}
\end{figure}
%%%%%%%%%%%%%%%%%%%%%%%%%%%%%%%%%%%%%%%%%%%%%%%%%%%%%%%%%

Functions $D^{(1)}(X)$ and $D^{(2)}(X)$ are usually called drift and diffusion coefficients respectively and they can be as simple as constants or linear functions of $X$, as e.g., the Ornstein-Uhlenbeck process, as well as more complicated nonlinear functions, typically polynomials up to a given order. In particular if $D^{(2)}$ is explicitly depending on $x$, the case is called multiplicative noise.

In all cases, through substitution of the selected functions into Equation~\ref{LangevinDisc} one is able to generate samples of series having the same statistical features and obeying the same dynamics.

Figure~\ref{fig01}a shows an illustration of a time series obtained through integration of Equation~\ref{LangevinDisc} for a cubic drift $D^{(1)}(X) = -X^3 + X$, and a quadratic diffusion, $D^{(2)}(X) = X^2 + 1$. Notice that these drift and diffusion coefficients describe a non-trivial dynamics, namely the underlying deterministic process, i.e., $D^{(2)} \equiv 0$, has two attractive fixed points at $X = \pm 1$. The processes tends to converge to one of two stable states being at the same time perturbed by a stochastic fluctuation ($D^{(2)} \neq 0$) which is able to push the system from one stable fixed point to the other. As shown in Figure~\ref{fig01}b, despite this non-trivial dynamics, the PDF is a Gaussian distribution with zero mean and unit standard deviation, the same PDF as for a simple Ornstein-Uhlenbeck process with $D^{(1)} = -X$ and $D^{(2)} = 1$.

This is one of many possible examples that illustrates the deep insight, which an evolution equation like in Equation~\ref{Langevin} can provide and which is not obtained by looking at a density distribution, see Appendix~\ref{append:twodyn} for further details.

\subsection{From stochastic data to the Langevin model}

As explained previously, it is easy to generate data through the integration of a stochastic equation, such as Equation~\ref{LangevinDisc}. More difficult is the inverse problem, to derive functions $D^{(1)}$ and $D^{(2)}$ from given data.

A condition to derive the drift and diffusion numerically is that the time-steps $\tau$ of the set of $X$-values are small enough (see \citet{Honisch2011} for details). If the system is at time $t$ in the state $x = X(t)$ the drift can be calculated for small $\tau$ by averaging over the difference, $X(t + \tau) - X(t)$, of the system state at $t + \tau$ and the state at $t$. Check Equation~\ref{LangevinDisc} above. This average is the first conditional moment of the series and it can be mathematically proven that its time derivative yields the drift coefficient. Similarly, computing the second conditional moment, i.e., the average squared differences between $X(t + \tau)$ and $x$, yields the diffusion coefficient \citep{Risken1996}. 

%%%%%%%%%%%%%%%%%%%%%%%%%%%%%%%%%%%%%%%%%%%%%%%%%%%%%%%%%%%
\begin{figure}
\centering
\includegraphics[width=\textwidth]{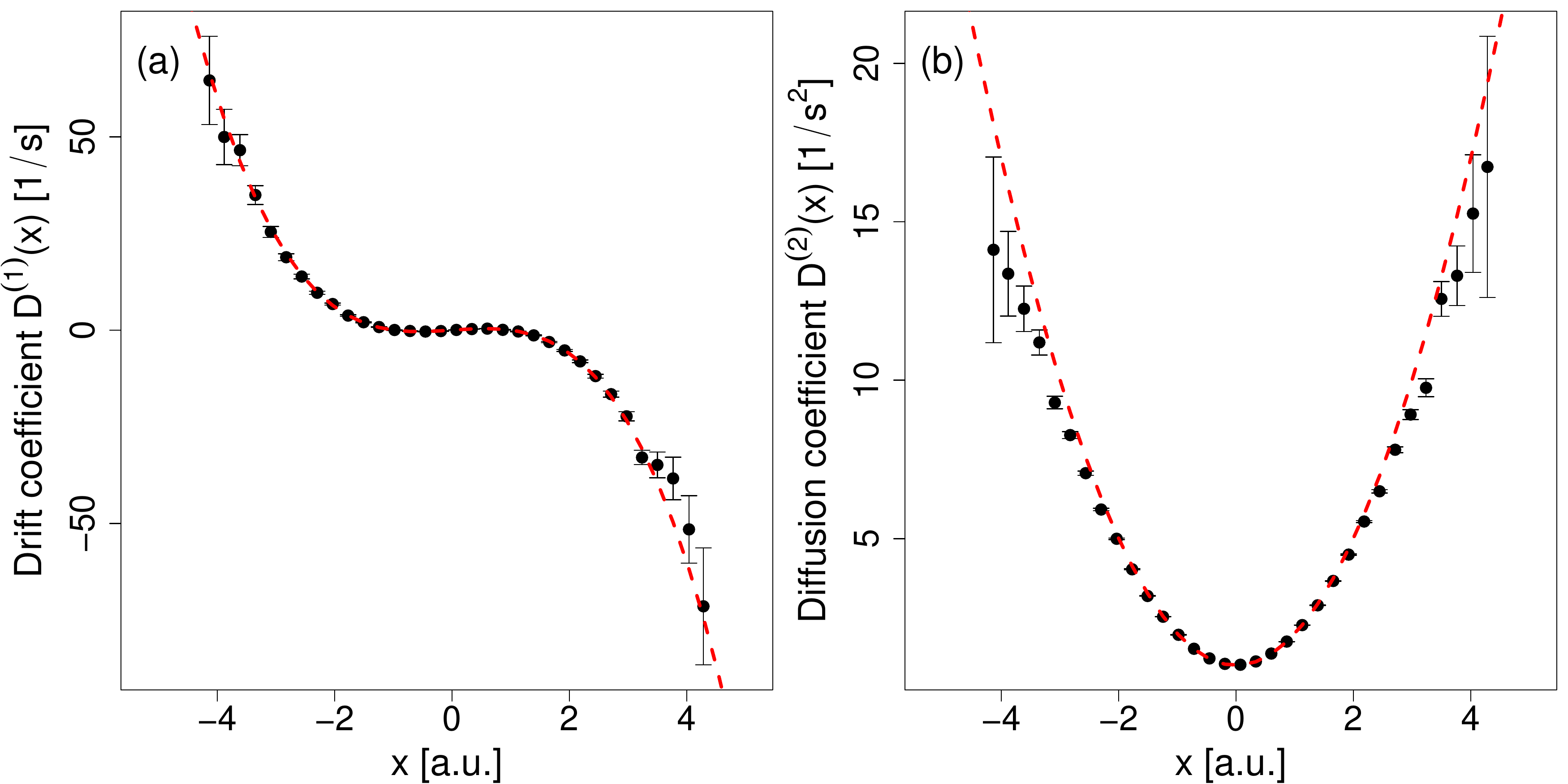}
\caption{One-dimensional Langevin Approach: {\bf (a)} drift coefficient, $D^{(1)}(x) = -x^3 + x$, and {\bf (b)} diffusion coefficient, $D^{(2)}(x) = x^2 + 1$. Circles indicate the numerical results while the red dashed line indicates the theoretical coefficient, used when generating the synthetic data. Here $10^7$ data points from the series illustrated in Figure~\ref{fig01}a, were used for computing the averaged conditional moments.}
\label{fig02}
\end{figure}
%%%%%%%%%%%%%%%%%%%%%%%%%%%%%%%%%%%%%%%%%%%%%%%%%%%%%%%%%%%

Therefore, having a series of data, one estimates the drift and diffusion by computing the averages of the first and second power of the differences between $X(t + \tau)$ and $x$:

\begin{equation} 
M^{(n)}(x,\tau) = \langle (X(t+\tau) - X(t))^n \rangle
|_{X(t) = x} ,
\label{condmom}
\end{equation} 

where $\langle\cdot\rangle$ represents the average over time $t$. Mathematically the drift and diffusion coefficients are defined as \citep{Risken1996}

\begin{equation} 
D^{(n)}(x) = \lim_{\tau \rightarrow 0} \frac{1}{n!\tau} 
M^{(n)}(x,\tau) ,
 \label{driftdiff}
\end{equation} 

which means that they are given by derivatives of the corresponding conditional moments $M^{(n)}(x, \tau)$ with respect to $\tau$. In many cases, for a fixed $x$, the conditional moments depend linearly on $\tau$ for the smallest range of $\tau$ values and consequently the drift and the diffusion coefficients at this state $x$ are estimated solely by the quotient between the corresponding conditional moment and
$\tau$ in this range. 

Figures~\ref{fig02}a and \ref{fig02}b show respectively the drift and diffusion coefficients of the series integrated in the previous section and sketched in Figure~\ref{fig01}. The theoretical expressions of the coefficients used when generating the synthetic data through integration of Equation~\ref{Langevin} are indicated as red dashed lines, while the estimated values of the coefficients at each selected bin are plotted with bullets.

\subsection{Back to the data}
\label{subsec:back}

The Langevin Approach summarized previously is applied under a few conditions, though, as we discus afterwards, when such conditions are not fulfilled in many cases it is still possible to overcome that and apply an alternative approach which also retrieves the dynamics underlying the
stochastic process. For the completeness of this paper and for the consistency of the application of our \proglang{R} functions, we advise the user to briefly test the data. Three conditions should in general be tested as a preliminary checking procedure and two further conditions can afterwards be tested as cross-checking.

The first condition is that the data series is stationary. Indeed, the averages for computing the conditional moments have to be taken over all $t=t_i$ where $X(t_i)=x$ (see Equation~\ref{condmom}). If the series is non-stationary these averages are in principle not meaningful.

The second condition is that the process should be Markovian, i.e., the present state should depend on the previous state solely. Mathematically it means an equivalence between two-point statistics, $p(X(t + \tau), X(t))$, and any higher-order statistics, $p(X(t + \tau), X(t), ..., X(t - n\tau))$. This equivalence leads to the following equality between conditional probabilities of finding a value of $X(t+\tau)$ under the condition that $X(t), X(t - \tau), ..., X(t - n\tau)$ have selected values:

\begin{equation}
p(X(t + \tau) \vert X(t)) = p(X(t + \tau) \vert X(t), ..., X(t - n\tau)) \, .
\label{conditional}
\end{equation}

This should hold for any positive integer $n$. In practice, one tests the equality for three-point statistics ($n = 1$) only and assumes that if the equality holds it will also hold for higher-order statistics, since all correlations shall decrease monotonically with time.

To test if the process is Markovian one can also use alternatively the Wilcoxon test \citep{Wilcoxon1945}, in case one is dealing with single variable stochastic processes. For details see \citet[Appendix A]{Renner2001}.

The third condition to be tested comes from a mathematical result called Pawula Theorem \citep{Risken1996}, from which it follows a
necessary condition for Equation~\ref{Langevin} to be valid: the fourth conditional moment must be constant, i.e., $D^{(4)} = 0$. To test that one computes its derivative with respect to the time-lag, the fourth coefficient

\begin{equation}
D^{(4)}(x) = \lim_{\tau \rightarrow 0} \frac{1}{4!\tau}\langle (X(t+\tau) - X(t))^4 \rangle |_{X(t) = x} 
 \label{D4}
\end{equation}

and check if it vanishes, i.e., if it is small compared to the diffusion coefficient: $D^{(4)}(x) \ll (D^{(2)}(x))^2 \; \forall x$. This coefficient is also useful for computing the numerical error of the diffusion coefficient \citep{lind2010}.

The tests whether conditions two and three hold ensure that $\Gamma(t)$ (see Equation~\ref{Langevin}) is $\delta$-correlated and Gaussian distributed.

If all these conditions are fulfilled the Langevin Approach can be carried out and the two functions, drift coefficient $D^{(1)}$ and diffusion coefficient $D^{(2)}$, can be derived from the given data. With the derived coefficients two additional cross-checking tests can be done.

The first one is to check if the stochastic force in Equation~\ref{Langevin} fulfills the two conditions of a $\delta$-correlated Gaussian white noise. To that end, one substitutes in Equation~\ref{LangevinDisc} the derived $D^{(1)}(X)$ and $D^{(2)}(X)$ and solves it with respect to $\eta(t)$:

\begin{equation}
\eta(t) = \frac{X(t + \tau) - X(t) - D^{(1)}(X)\tau}{\sqrt{D^{(2)}(X)\tau}} \, .
\end{equation}

Taking $\tau$ as the time-step of the observed time-series and substituting in $X(t + \tau)$ and $X(t)$ successive values of that series one re-obtains a series for $\eta(t)$ which should be normally distributed. 

The second cross-checking test is to substitute in Equation~\ref{LangevinDisc} the derived $D^{(1)}(X)$ and $D^{(2)}(X)$ coefficients, generate synthetic series and compare if its increments

\begin{equation}
\Delta_{\tau}(t) = X(t + \tau) - X(t)
\label{increments}
\end{equation}

have the same distribution as the original series for a fixed $\tau$ spanning from the time-step of the original series up to two or more orders of magnitude larger.

Some extra care should be taken if the derived $D^{(1)}(X)$ and $D^{(2)}(X)$ coefficients show linear drift and quadratic diffusion forms as this is also the case for every Langevin process if the sampling interval is large compared to the relaxation time of the process. \citet{Riera2010} presented a method for cross-checking in this case.

Notice that, though the fulfillment of all such conditions through the proposed preliminary tests and cross-checking tests guarantees that the Langevin Approach can be applied, the rejection of one or more of these tests is still no reason for avoiding this approach. In Section~\ref{sec:remarks} we will come back to this issue.

\section{Implementation and examples}
\label{sec:implementation}

In this section we present the implementation of the Langevin Approach describing the two available \proglang{R} functions, \code{Langevin1D} and \code{Langevin2D}. The function \code{Langevin1D} deals with single time-series while \code{Langevin2D} should be used for two-dimensional cases, when one has two stochastic variables to be analyzed simultaneously.

The one-dimensional case deals with an evolution equation similar to Equation~\ref{Langevin} and the two-dimensional case comprehends two stochastic variables, $X_1(t)$ and $X_2(t)$, governed by:

\begin{equation}
\frac{d}{dt}
\left [
\begin{array}{c} 
X_1 \\
X_2 
\end{array}
\right ]
 = 
\left [
\begin{array}{c}
D^{(1)}_1(X_1,X_2) \\
D^{(1)}_2(X_1,X_2) 
\end{array}
\right ]
+ \left [
\begin{array}{cc} 
g_{11}(X_1,X_2) &g_{12}(X_1,X_2) \\
g_{21}(X_1,X_2) &g_{22}(X_1,X_2) 
\end{array}
\right ]
\left [
\begin{array}{cc}
\Gamma_1(t) \\
\Gamma_2(t) 
\end{array}
\right ]
\label{Langevin2D}
\end{equation} 

where clearly now the drift function $\mathbf{D}^{(1)} = (D^{(1)}_1, D^{(1)}_2)$ is a two-dimensional vector and the diffusion coefficient is a $2\times 2$-matrix given by $\mathbf{D}^{(2)} = \mathbf{g}\mathbf{g}^T$, i.e., $D^{(2)}_{ij} = \sum_k g_{ik}g_{jk}$. Similar to the one-dimensional case the integration of Equation~\ref{Langevin2D} follows from a simple Euler scheme leading to:

\begin{equation}
\begin{array}{r@{}l}
\left [
\begin{array}{c} 
X_1(t+\tau) \\ X_2(t+\tau)
\end{array}
\right ]
 = 
\left [
\begin{array}{c} 
X_1(t) \\ X_2(t)
\end{array}
\right ]
&{}+ \tau\left [
\begin{array}{c}
D^{(1)}_1(X_1,X_2) \\ D^{(1)}_2(X_1,X_2) 
\end{array}
\right ] \\
&{}+ \sqrt{\tau}\left [
\begin{array}{cc} 
g_{11}(X_1,X_2) &g_{12}(X_1,X_2) \\
g_{21}(X_1,X_2) &g_{22}(X_1,X_2) 
\end{array}
\right ]
\left [
\begin{array}{cc}
\eta_1(t) \\
\eta_2(t) 
\end{array}
\right ]
\end{array}
\label{Langevin2DDisc}
\end{equation} 

where $\eta_1(t)$ and $\eta_2(t)$ are two independent normally distributed random variables.

In our implementation the conditional moments $M^{(n)}(x,\tau)$, Equation~\ref{condmom}, are estimated by dividing the state space of $x$ in $N$ intervals, or bins, $(I_1, ..., I_N)$ and calculating the mean values for each interval $I_i$:

\begin{equation}
  M^{(n)}(x,\tau) = \langle (X(t+\tau) - X(t))^n \rangle |_{X(t) \in I_i}\,.
  \label{condmomI}
\end{equation}

For estimating the drift and diffusion coefficients from the conditional moments we insert Equation~\ref{LangevinDisc} into Equation~\ref{condmomI} and apply the conditional averages for $n = 1, 2$ leading to:

\begin{eqnarray}
  \label{finiteTauApproxM1}
  M^{(1)}(x,\tau) &\approx& D^{(1)}(x)\tau\,,\\
  M^{(2)}(x,\tau) &\approx& 2D^{(2)}(x)\tau + (D^{(1)}(x)\tau)^2 \,.
  \label{finiteTauApproxM2}
\end{eqnarray}

Important to notice is that for $M^{(2)}(x,\tau)$, Equation~\ref{finiteTauApproxM2}, a term quadratic in $D^{(1)}(x)$ and $\tau$ has to be considered. We estimate drift and diffusion coefficients from the slope of a weighted linear regression of Equations~\ref{finiteTauApproxM1} and \ref{finiteTauApproxM2}.

The implementation of the functions heavily relies on the \proglang{C++} linear algebra library {\tt Armadillo} \citep{Sanderson2010} for which \pkg{RcppArmadillo} and \pkg{Rcpp} provide the integration with \proglang{R} \citep{Eddelbuettel2011,Eddelbuettel2014}. We choose {\tt Armadillo} as it results in fast code especially for large data sets and has an easy readable syntax.
The functions \code{Langevin1D} and \code{Langevin2D} use {\tt OpenMP} \citep{Dagum1998} if available to take advantage of shared memory multiprocessing. Here we parallelize the evaluation of the drift and diffusion coefficients for the bins as their evaluation is independent for each bin.

In the following subsections we present one- and two-dimensional examples of Langevin processes and walk through the analysis based on the framework described in the previous section.

\subsection[Langevin1D: analyzing one-dimensional data sets]{\code{Langevin1D}: analyzing one-dimensional data sets}
\label{subsec:Langevin1D}

As an example we integrate the Langevin equation illustrated in Figure~\ref{fig01}a with cubic drift and quadratic diffusion, namely

\begin{equation} 
\frac{dx}{dt} = x(t) - x^3(t) + \sqrt{x^2(t) + 1} \, \Gamma(t) \, .
\label{ex1D}
\end{equation}

The presented package provides the function \code{timeseries1D} to do the integration using an Euler integration scheme:

\begin{CodeChunk}
\begin{CodeInput}
R> library("Langevin")
R> sf <- 1000
R> set.seed(4711)
R> x <- timeseries1D(N = 1e7, d11 = 1, d13 = -1, d22 = 1, d20 = 1, sf = sf)
\end{CodeInput}
\end{CodeChunk}

Extracting drift and diffusion coefficients from the generated time series is done by the function \code{Langevin1D}. Here two parameters that are important for the estimation have to be given as arguments.

The first one is the number of \code{bins} dividing the variable space $x$ in discrete bins at which drift and diffusion are estimated. This integer should not be so large that each bin does no longer include a significant number of points (typically $\sim 100$) and also not so small that no dependence of the drift and diffusion on the state variable can be observed.

The second parameter is the vector \code{steps} to calculate the conditional moments for different $\tau$ values (Equation~\ref{condmom}). The conditional moments will be computed for each bin and for each step. For each bin, a linear fit is computed for all steps in \code{steps}. Typically a vector of up to ten steps is given in samples ($=\tau \cdot \mathrm{sf}$).

\begin{CodeChunk}
\begin{CodeInput}
R> bins <- 40
R> steps <- c(1:3)

R> ests <- Langevin1D(x, bins, steps)
\end{CodeInput}
\end{CodeChunk}

From the resulting list \code{ests}, plots of the estimated drift and diffusion coefficients can be generated (see Figure~\ref{fig02}). Here we use \pkg{plotrix} \citep{Lemon2006} to add errorbars.

\begin{CodeChunk}
\begin{CodeInput}
R> library("plotrix")
R> attach(ests)
R> par(mfrow = c(1, 2))
R> plotCI(mean_bin, D1, uiw = eD1, xlab = "x [a.u.]", 
+    ylab = expression(paste("Drift coefficient ", D^(1), "(x) [a.u.]")), 
+    cex = 2, pch = 20)
R> lines(mean_bin, mean_bin - mean_bin^3, col = "red", lwd = 3, lty = 2)
R> plotCI(mean_bin, D2, uiw = eD2, xlab = "x [a.u.]", 
+    ylab = expression(paste("Diffusion coefficient ", D^(2), "(x) [a.u.]")),
+    cex = 2, pch = 20)
R> lines(mean_bin, mean_bin^2 + 1, col = "red", lwd = 3, lty = 2)
\end{CodeInput}
\end{CodeChunk}

We now want to walk through some of the remarks given in Section~\ref{subsec:back} to check if the conditions under which we applied the Langevin Approach are fulfilled. We do not check if the time series is stationary and fulfills the Markovian properties, since here we already know this (as we are using synthetic data).

Therefore we concentrate on cross-checking the estimated drift and diffusion coefficients. For checking if $D^{(4)}(X)$ is small compared to $D^{(2)}(X)$ (Pawula Theorem) we use the function \code{summary} which also computes the ratio between $D^{(4)}$ and $(D^{(2)})^2$:

\begin{CodeChunk}
\begin{CodeInput}
R> summary(ests)
\end{CodeInput}
\begin{CodeOutput}
 Number of bins: 40
 Population of the bins:
	Min.  : 3
	Median: 32034
	Mean  : 250000
	Max.  : 1053446
 Number of NA's for D1: 7
 Number of NA's for D2: 7
 Ratio between D4 and D2^2:
	Min.  : 0.002004
	Median: 0.002102
	Mean  : 0.002385
	Max.  : 0.004487
\end{CodeOutput}
\end{CodeChunk}

The result shows that $D^{(4)}(X)$ is smaller than $0.5\%$ of the squared diffusion coefficient, indicating the necessary condition of the Pawula Theorem holds.

As a second cross-check we compare the increments, as defined in Equation~\ref{increments}, of the original time series with the ones computed from the reconstructed time series based on the estimated drift and diffusion functions. 

To this end we fit a cubic function to the estimated drift coefficient and a quadratic function to the diffusion coefficient:

\begin{CodeChunk}
\begin{CodeInput}
R> estD1 <- coef(lm(D1 ~ mean_bin + I(mean_bin^2) + I(mean_bin^3), weights = 1/eD1))
R> estD2 <- coef(lm(D2 ~ mean_bin + I(mean_bin^2), weights = 1/eD2))
\end{CodeInput}
\end{CodeChunk}

The resulting coefficients are used to generate a new time series with \code{timeseries1D}:

\begin{CodeChunk}
\begin{CodeInput}
R> rec_x <- timeseries1D(N = 1e7, d10 = estD1[1], d11 = estD1[2], d12 = estD1[3],
+    d13 = estD1[4], d20 = estD2[1], d21 = estD2[2], d22 = estD2[3], sf = sf)
\end{CodeInput}
\end{CodeChunk}

We want to emphasize here that the Langevin Approach does not require the drift and the diffusion coefficients to be of any particular functional form, from the estimated coefficients one could directly integrate a stochastic time series which can be used to calculate the increments. We fit the estimated coefficients to polynomials only to be able to use the function \code{timeseries1D} for the integration.

From the original and the reconstructed time series we now calculate PDFs of the increments for different $\tau$ and plot them to inspect their agreement visually:

\begin{CodeChunk}
\begin{CodeInput}
R> plot(1,1, log = "y", type = "n", xlim = c(-11, 12), ylim = c(1e-17, 5), 
+    xlab = expression(Delta[tau]/sigma[Delta[tau]]), ylab = "density")

R> tau <- c(1,10,100,1000)
R> for(i in 1:4) {
+      delta <- diff(Ux, lag = tau[i])
+      rec_delta <- diff(rec_x, lag = tau[i])
+      den <- density(delta)
+      den$x <- den$x/sd(delta, na.rm = TRUE)
+      rec_den <- density(rec_delta)
+      rec_den$x <- rec_den$x/sd(rec_delta, na.rm = TRUE)
+      lines(den, lwd = 2, col = i)
+      lines(rec_den, lwd=2, lty = 2, col = i)
+  }
\end{CodeInput}
\end{CodeChunk}

\begin{figure}
\centering
\includegraphics[width=\textwidth]{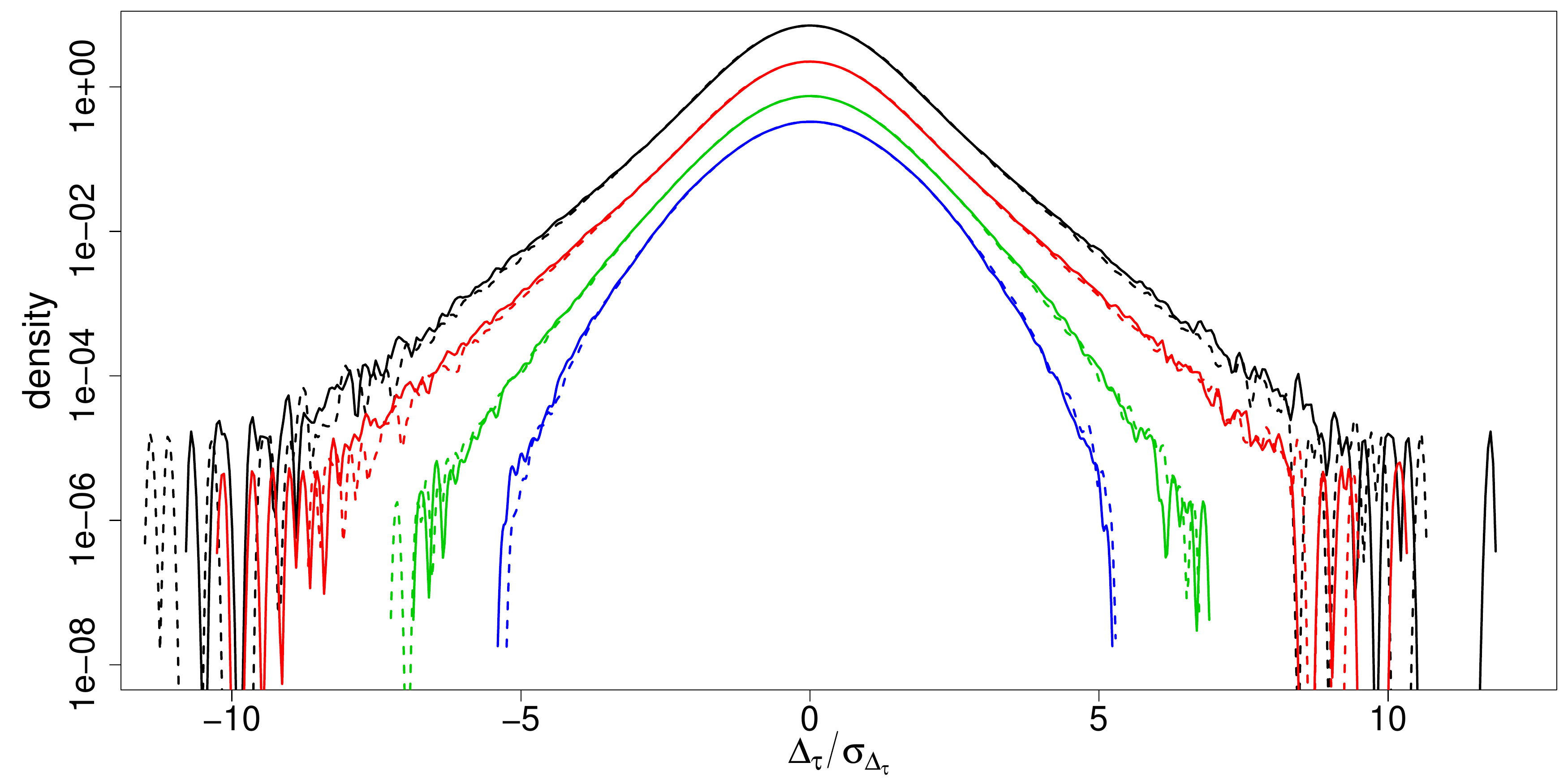}
\caption{PDFs of the increments for $\tau = 1$, $\tau = 10$, $\tau = 100$ and $\tau = 1000$ time lags (from top to bottom). Solid lines show the results for the original time series, broken lines the result for the reconstructed time series.}
\label{fig02rec}
\end{figure}

Figure~\ref{fig02rec} shows the output: there is indeed good agreement of both increment PDFs for a wide range of $\tau$ values. Therefore we can assume that our estimated drift and diffusion coefficients describe the process sufficiently.

%%%%%%%%%%%%%%%%%%%%%%%%%%%%%%%%%%%%%%%%%%%%%%%%%%%%%%%%%%%
%%%%%%%%%%%%%%%%%%%%%%%%%%%%%%%%%%%%%%%%%%%%%%%%%%%%%%%%%%%

Notice once again that while the PDF of the series generated by Equation~\ref{ex1D} is the same as the one of the simple Ornstein-Uhlenbeck process, $\frac{dx}{dt} = -x(t) + \Gamma(t)$, our Langevin Approach is able to uncover the correct dynamics with a bistable drift and a non-constant diffusion. See Appendix~\ref{append:twodyn}.

\subsection[Langevin2D: analyzing two-dimensional data sets]{\code{Langevin2D}: analyzing two-dimensional data sets}

\begin{figure}
\centering
\includegraphics[width=\textwidth]{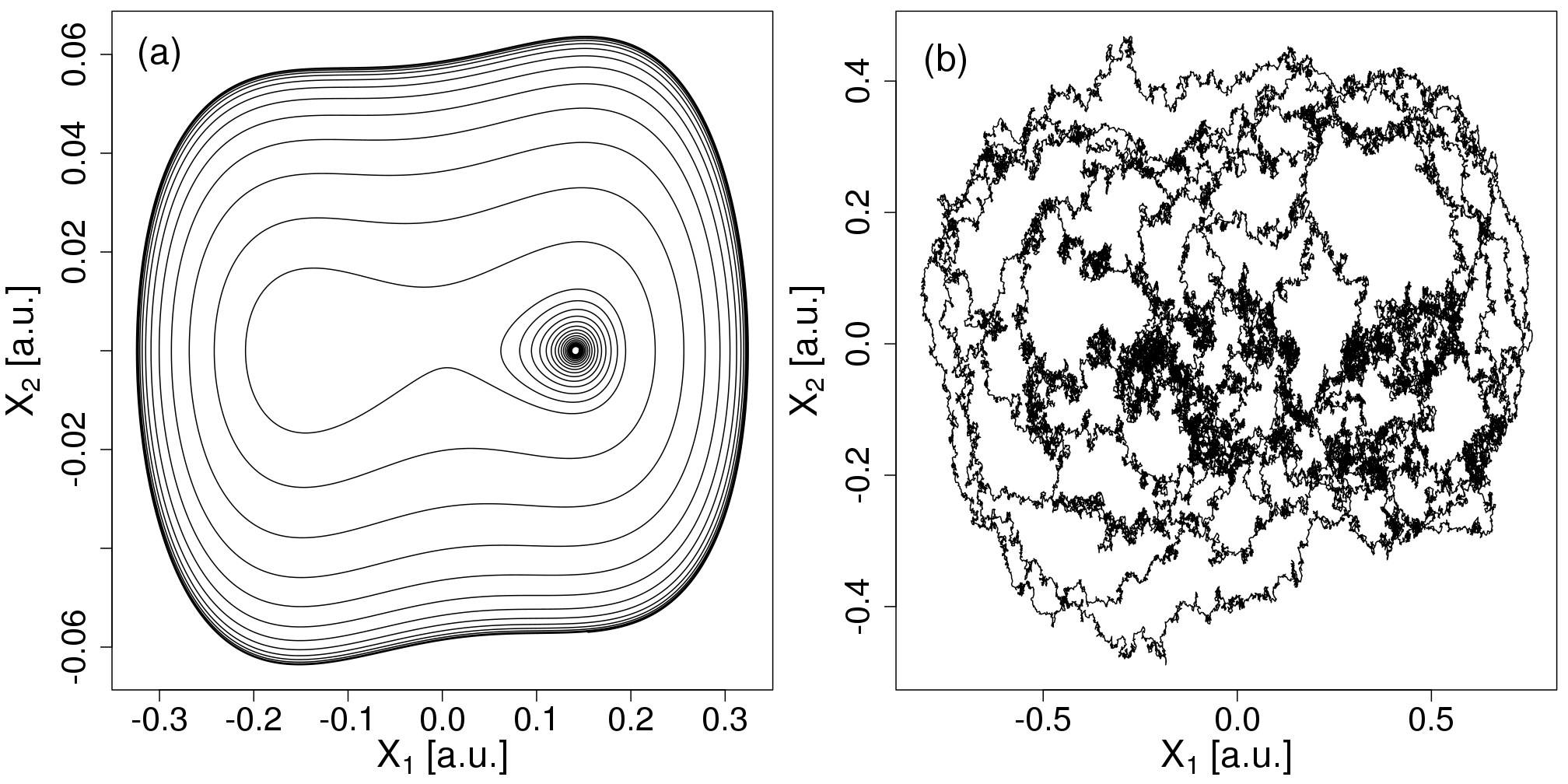}
\caption{{\bf (a)} Trajectory $(X_1(t), X_2(t))$ from Equations~\ref{ex2D} with $a = 0$ and {\bf (b)} the same trajectory integrating the same equations with non-zero stochastic terms ($a = 0.05$). For plotting $10^6$ resp. $10^5$ data points where used.} 
\label{fig03}
\end{figure}

As a two-dimensional example we integrate the coupled Langevin equations in Equations~\ref{Langevin2D} for a particular choice of the drift and diffusion coefficients, namely \citep{Siegert1998}

\begin{subequations}
\begin{eqnarray} 
\frac{X_1}{dt} &=& X_2 + a\Gamma_1(t) \\
\frac{X_2}{dt} &=& 0.02X_1 + 0.03X_2 - X_1^3 - X_1^2X_2 + a\Gamma_2(t) \, ,
\end{eqnarray}
\label{ex2D}
\end{subequations}

where $a$ is a constant. Figure~\ref{fig03}a shows the integrated trajectory $(X_1, X_2)$ for $a = 0$, a case where no stochastic contribution is present, whereas in Figure~\ref{fig03}b the same trajectory is plotted now with stochastic forces having a constant amplitude of $a = 0.05$. 

%%%%%%%%%%%%%%%%%%%%%%%%%%%%%%%%%%%%%%%%%%%%%%%%%%%%%%%%%
%%%%%%%%%%%%%%%%%%%%%%%%%%%%%%%%%%%%%%%%%%%%%%%%%%%%%%%%%

The integration is performed by \code{timeseries2D}. Drift and diffusion functions are full cubic and quadratic polynomials respectively and the elements $a_{ij}$ of the matrices are defined by the corresponding equations for the drift and diffusion terms (see Equations~\ref{Langevin2D} and \ref{Langevin2DDisc}):

$$ D^{(1)}_{1,2} = \sum_{i=1}^4 \sum_{j=1}^{5-i} a_{ij} x_1^{(i-1)}x_2^{(j-1)} \quad \mathrm{and} \quad g_{11,12,21,22} = \sum_{i=1}^3 \sum_{j=1}^{4-i} a_{ij} x_1^{(i-1)}x_2^{(j-1)} \; .$$

Estimating the drift and diffusion coefficients is done by \code{Langevin2D}, here the same rules for \code{bins} and \code{steps} apply as for the one-dimensional case.

\begin{figure}
\centering
\includegraphics[width=\textwidth]{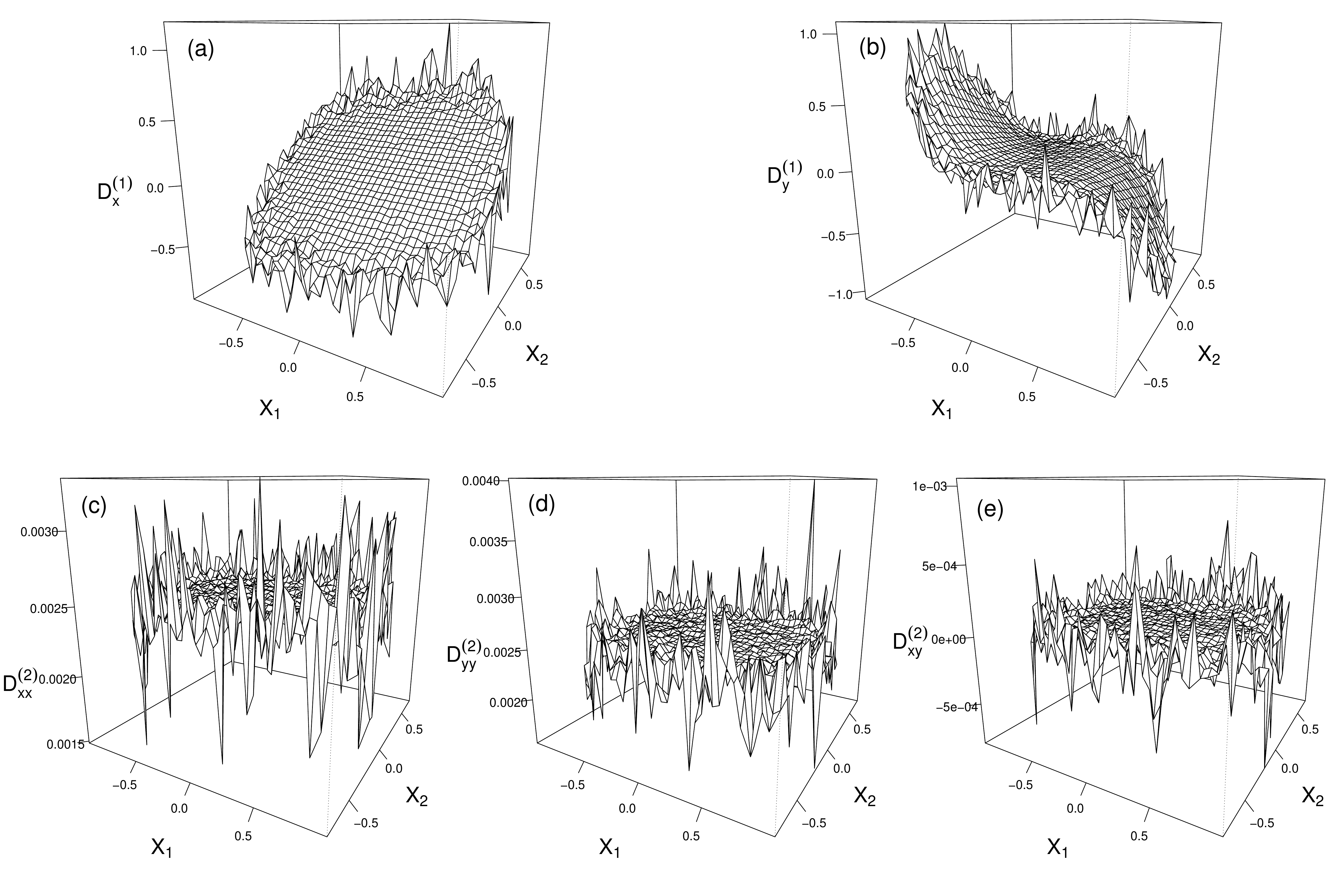}
\caption{Drift coefficient of {\bf (a)} the $X_1$ component, $D^{(1)}_1$, and {\bf (b)} the $X_2$ component, $D^{(1)}_2$, together with all diffusion coefficients, namely {\bf (c)} $D^{(2)}_{11}$, {\bf (d)} $D^{(2)}_{22}$, {\bf (e)} $D^{(2)}_{12} = D^{(2)}_{21}$. See Equation~\ref{Langevin2D}. Estimated with added noise, i.e., $a = 0.05$ in Equation~\ref{ex2D}.}
\label{fig04}
\end{figure}

The results shown in Figure~\ref{fig04} are generated by the following command lines:

\begin{CodeChunk}
\begin{CodeInput}
R> D1_1 <- matrix(0, nrow = 4, ncol = 4)
R> D1_1[1, 2] <- 1
R> D1_2 <- matrix(0, nrow = 4, ncol = 4)
R> D1_2[2, 1] <- 0.02
R> D1_2[1, 2] <- 0.03
R> D1_2[4, 1] <- -1
R> D1_2[3, 2] <- -1

R> g_11 <- matrix(0, nrow = 3, ncol = 3)
R> g_12 <- matrix(0, nrow = 3, ncol = 3)
R> g_21 <- matrix(0, nrow = 3, ncol = 3)
R> g_22 <- matrix(0, nrow = 3, ncol = 3)
R> g_11[1, 1] <- 0.0025
R> g_22[1, 1] <- 0.0025

R> set.seed(4711)
R> x <- timeseries2D(N = 1e8, 0.145, 0.0002, D1_1, D1_2,
+    g_11, g_12, g_21, g_22, sf = sf)

R> ests <- Langevin2D(x, bins, steps)
\end{CodeInput}
\end{CodeChunk}

%%%%%%%%%%%%%%%%%%%%%%%%%%%%%%%%%%%%%%%%%%%%%%%%%%%%%%%%%
%%%%%%%%%%%%%%%%%%%%%%%%%%%%%%%%%%%%%%%%%%%%%%%%%%%%%%%%%

The numerical results can be properly fitted through the functions used for the integration in Equations~\ref{ex2D}, namely: $D^{(1)}_1 = X_2$, $D^{(1)}_2 = 0.02X_1 + 0.03X_2 - X_1^3 - X_1^2X_2$, $D^{(2)}_{11} = D^{(2)}_{22} = 0.05^2$ and $D^{(2)}_{12} = D^{(2)}_{21} = 0$. Notice that the large deviations in the boundaries are due to the finite length of the time series and thus the lower population in the boundary bins resulting in a poorer estimation of the drift and the diffusion.

\section[A glimpse beyond the Langevin package]{A glimpse beyond the \pkg{Langevin} package}
\label{sec:remarks}

The two examples exposed above show cases where all conditions are fulfilled. When analyzing real empirical data sets this is often not the case: one or more of the conditions under which the Langevin Approach is applied are not met. Still, in the last years we developed different alternatives and extensions to this approach to overcome specific situations in stochastic data analysis. In this section we briefly describe these alternatives and extensions.

One first problem that researchers face is the non-stationary character often appearing in real data. Here, one of two approaches may be possible. One is to ascertain if for ``shorter'' time-windows of the data series stationarity may be assumed. In case the data set can be decomposed in a series of time-windows which may overlap, each one having more or less constant statistical moments of the observable, the Langevin approach can be applied separately to each one of them, yielding a set of drift and diffusion coefficients, one for each time-window. In the end one extracts one drift and one diffusion coefficient, both functions of the observable and also of time.

Another possibility to handle non-stationary data sets is to check if they can be conditioned to other observables. In that case, considering the periods of the data sets associated to a particular value of the conditioning observable may be itself stationary. This is the case of the stochastic series measured of wind turbines \citep{Waechter2011,Lind2014}. The power output of one wind turbine or the loads applied to it by the wind field are two observables whose measurement series are by themselves non-stationary. The wind velocity is the observable driving those properties and it is also non-stationary. However, we have shown that both wind power production \citep{Waechter2011} and instantaneous loads \citep{Lind2014} can be analyzed through the Langevin Approach if we conditioned both the drift and the diffusion coefficients to each particular admissible value of the wind speed.

The second condition listed above is the Markov property. When the series of measurements fails to fulfill the Markov tests described above, it cannot be reconstructed through stochastic Euler integration since the next state cannot be estimated from the present state alone (see Equation~\ref{LangevinDisc}). This happens, for instance, when a Markov process is spoiled by additional additive noise when a measurement is taken \citep[see][]{Kleinhans2007d}. While the process alone, $X(t)$, is Markovian, the actual measurement, which retrieves $X(t) + Y(t)$, does not fulfill the Markov property. In such cases the limits computed for the coefficients $D^{(1)}$ and $D^{(2)}$ diverge (see Equation~\ref{driftdiff}): when $\tau \to 0$ the conditional moment for the measured values (numerator) does not vanish. Still, it is frequently possible to obtain the correct drift and diffusion coefficients for the Markov process $X(t)$ through simple changes of their estimates \citep{Boettcher2006,Lehle2011,Lehle2013,Scholz2015}. In cases of correlated noise where $\langle \tilde{\Gamma}(t) \rangle = 0$ holds the drift coefficient $D^{(1)}(X)$ can still be reconstructed correctly.

A third problem that may appear during preliminary tests of empirical data is the non-vanishing fourth coefficient $D^{(4)}$. As stated above in Section~\ref{subsec:back}, according to Pawula Theorem \citep{Risken1996} the fourth conditional moment must be independent of the time-lag $\tau$. If not, one cannot assume that the stochastic process is governed by a Langevin equation, Equation~\ref{Langevin}. However, in such cases, although no evolution equation can be extracted and therefore the estimated functions $D^{(1)}$ and $D^{(2)}$ have not the meaning of drift and diffusion contributions, one can still use both to provide valuable insight about the system being analyzed. One example is the work of \citet{Rinn2012} on in-situ analysis of the elastic features of a mechanical beam structure for realistic excitations with correlated noise as it appears in real-world situations. They could show that the slope of the drift coefficient $D^{(1)}$ is a sensitive indicator of the damage and compared to frequency based approaches, like power spectra, which estimate changes of the eigenfrequency of the structure, it is even more sensitive to small damages.

Finally, it is also important to stress that, while the functions of the presented package were prepared for analyzing data series as processes in time, the Langevin Approach can be adapted for analyzing processes in scale. In fact, when the process is not Markovian in time, violating Equation~\ref{conditional}, there is the possibility that it is Markovian in ``scale''. What does this mean? It means that the increments $\Delta_{\tau}$ introduced above follow a Markovian process in $\tau$ i.e., in time-lags but are instationary. Such analysis in scale is able to reproduce e.g., turbulence energy cascades \citep{Friedrich1997, Stresing2010} or ocean rogue waves \citep{Hadjihosseini2014}.

More details on all these extensions and alternatives to the Langevin Approach can be found in \citet{Friedrich2011}.

\section{Discussion and conclusions}
\label{sec:conclusions}

In this paper we present an \proglang{R} package for stochastic data analysis that is able to extract the stochastic evolution equations of physical properties from sets of their measurements.

The introduced functions serve as a framework to analyze one- and two-dimensional time series. They provide estimation of drift and diffusion coefficients describing the deterministic and the stochastic part of the analyzed process respectively. Integrating Langevin processes numerically enables one for cross-checking the obtained result and for generation of synthetic data sets.

Through illustrative examples we have shown that the Langevin evolution equation is able to uncover complex dynamics, even in cases when the associated statistics is identical to many other stochastic processes. 

The presented package can be straightforwardly applied by \proglang{R}-users and it implies only a few preliminary tests to ascertain if all conditions on which the Langevin Approach is built are fulfilled. In case they are not, we briefly explain how to overcome them with simple extensions to the method that were already successfully applied in several applications \citep{Friedrich2011}.

Still, additional improvements of the presented functions are possible. For instance, instead of using the common average bin value when performing the binning of the data, one can incorporate a kernel-based regression of such values \citep{Lamouroux2009} or a maximum likelihood framework \citep{Kleinhans2012} for estimating the drift and diffusion functions. Moreover, the approach can also be applied to data having low sampling rates \citep{Honisch2011}.

\section*{Acknowledgments}

The authors thank Constantino Garcia Mart\'inez (University of Santiago de Compostela, Spain) for useful discussions about our methods and motivating in sharing their implementation with a broader audience during his visit to our group. PGL thanks the German Environment Ministry as part of the research project ``Probabilistic loads description, monitoring, and reduction for the next generation offshore wind turbines (OWEA Loads)'' under grant number 0325577B.

\appendix
\section{Different stochastic dynamics, same stationary distribution}
\label{append:twodyn}

In this appendix we show that a large family of two-point statistical distributions, each one univocaly defining one Langevin equation, Equation~\ref{Langevin}, corresponds to a one-point statistics given by the standard normal distribution

\begin{equation} 
P_0(X)\propto \exp{\left ( -\frac{X^2}{2} \right )} ,
\label{normal}
\end{equation} 

i.e., a Gaussian distribution with zero mean and unit variance.

To that end, we start with one important remark concerning the evolution equation of one stochastic variable $X$, Equation~\ref{Langevin}: this equation is related to an another evolution equation, namely the one of the probability density function (PDF) of $X$, so-called Fokker-Planck Equation \citep{Risken1996}:

\begin{equation} 
\frac{\partial P(X)}{\partial t} =
\left ( -\frac{\partial }{\partial X} D^{(1)}(X) + \frac{\partial^2 }{\partial
  X^2} D^{(2)}(X) \right ) P(X) \, .
\label{FP}
\end{equation} 

The stationary solution ($\frac{\partial P}{\partial t} = 0$) of the one-dimensional Fokker-Planck is given by \citet{Risken1996}:

\begin{equation}
P(X)\propto \frac{1}{D^{(2)}(X)} \exp{\left ( \int_X
      \frac{D^{(1)}(x)}{D^{(2)}(x)}dx \right )} \, .
\label{stationary}
\end{equation}

For the simple Ornstein-Uhlenbeck process, governed by Equation~\ref{Langevin} with $D^{(1)} = -x$ and $D^{(2)} = 1$, the stationary PDF reduces to $P_0$ in Equation~\ref{normal}.

One could, however, consider a much more complex dynamics such as the one exemplified in this paper, with a bistable (cubic) drift coefficient and a non-constant diffusion, depending quadratically on the stochastic variable $X$:

\begin{subequations} 
\begin{eqnarray} 
D^{(1)}(X) &=& aX(b - X)(b + X) \, ,\\
D^{(2)}(X) &=& c + dX^2 \, .
\end{eqnarray}
\label{thisol}
\end{subequations}

Here, $D^{(1)}$ has two stable fixed points at $\pm b$, with a maximum amplitude between them proportional to $a$, while $D^{(2)}$ has a minimum value $c$ and a broadness proportional to $1/d$.

Substituting the cubic drift and the quadratic diffusion, given in Equations~\ref{thisol}, into the stationary solution, Equation~\ref{stationary}, and integrating, yields the stationary solution:

\begin{eqnarray} 
P(X) &\propto& \frac{1}{c+dX^2} \exp{\left ( \int_X
                         \frac{ab^2x-ax^3}{c+dx^2}dx \right )}  \cr
  & & \cr
       &=& \frac{1}{d} \left ( \frac{c}{d} +X^2 \right
       )^{\frac{a}{2d}(b^2+\frac{c}{d})-1} \exp{\left (   -
           \frac{X^2}{2\frac{d}{a}} \right )} .
\label{complexstationary}
 \end{eqnarray}

As one sees, the solution in Equation~\ref{complexstationary} has, in general, not only a Gaussian part, like Ornstein-Uhlenbeck processes, but also a polynomial part with an exponent depending on all parameters of $D^{(1)}$ and $D^{(2)}$. However, if the exponent is exactly zero, 

\begin{equation}
\frac{a}{2d}\left ( b^2+\frac{c}{d} \right ) - 1 = 0
\label{exponent}
\end{equation}

the polynomial part vanishes and the stationary solution reduces to the Gaussian distribution. In the example used in Section~\ref{subsec:Langevin1D} with $a = b = c = d = 1$, this is the case (see Figures~\ref{fig01} and \ref{fig02} and Equation~\ref{ex1D}).

In general, the one-point statistic in the stationary regime given by Equation~\ref{stationary} yields Gaussian distributions even in more complex dynamics than the one here chosen. One only needs to have a drift coefficient given by one polynomial of odd degree $n > 0$ and simultaneously have a diffusion coefficient given by a polynomial of degree $n-1$. In that case, whatever general expression both coefficients have, it is always possible to find a combination of their parameter values for which the quotient $D^{(1)}/D^{(2)}$ in the stationary solution reduces to a linear function in $x$ yielding the PDF of a Gaussian distribution. The two-point statistic, $P(X(t)\vert X(t-\tau))$, however is able to distinguish between sets of $(D^{(1)}, D^{(2)})$ yielding the same one-point statistic.

The ambiguity of one-point statistics in characterizing the dynamics of stochastic processes in general, motivates the Langevin Approach implemented in our \proglang{R} package. Our approach has the advantage of being parameter free: since it computes numerically $D^{(1)}$ and $D^{(2)}$ without any given Ansatz, it can easily distinguish between higher-order drift and diffusion coefficients.

\bibliography{Langevin-package}

\begin{thebibliography}{30}
\newcommand{\enquote}[1]{``#1''}
\providecommand{\natexlab}[1]{#1}
\providecommand{\url}[1]{\texttt{#1}}
\providecommand{\urlprefix}{URL }
\expandafter\ifx\csname urlstyle\endcsname\relax
  \providecommand{\doi}[1]{doi:\discretionary{}{}{}#1}\else
  \providecommand{\doi}{doi:\discretionary{}{}{}\begingroup
  \urlstyle{rm}\Url}\fi
\providecommand{\eprint}[2][]{\url{#2}}

\bibitem[{B\"ottcher \emph{et~al.}(2006)B\"ottcher, Peinke, Kleinhans,
  Friedrich, Lind, and Haase}]{Boettcher2006}
B\"ottcher F, Peinke J, Kleinhans D, Friedrich R, Lind PG, Haase M (2006).
\newblock \enquote{Reconstruction of Complex Dynamical Systems Affected by
  Strong Measrement Noise.}
\newblock \emph{Physical Review Letters}, \textbf{97}, 090603.
\newblock \doi{10.1103/PhysRevLett.97.090603}.

\bibitem[{Dagum and Menon(1998)}]{Dagum1998}
Dagum L, Menon R (1998).
\newblock \enquote{OpenMP: An Industry Standard {API} for Shared-memory
  Programming.}
\newblock \emph{Computational Science \& Engineering, IEEE}, \textbf{5}(1),
  46--55.
\newblock \doi{10.1109/99.660313}.

\bibitem[{Eddelbuettel and Fran\c{c}ois(2011)}]{Eddelbuettel2011}
Eddelbuettel D, Fran\c{c}ois R (2011).
\newblock \enquote{\pkg{Rcpp}: Seamless \proglang{R} and \proglang{C++}
  Integration.}
\newblock \emph{Journal of Statistical Software}, \textbf{40}(8), 1--18.
\newblock \urlprefix\url{http://www.jstatsoft.org/v40/i08/}.

\bibitem[{Eddelbuettel and Sanderson(2014)}]{Eddelbuettel2014}
Eddelbuettel D, Sanderson C (2014).
\newblock \enquote{\pkg{RcppArmadillo}: Accelerating \proglang{R} with
  High-performance \proglang{C++} Linear Algebra.}
\newblock \emph{Computational Statistics and Data Analysis}, \textbf{71},
  1054--1063.
\newblock \doi{10.1016/j.csda.2013.02.005}.

\bibitem[{Friedrich and Peinke(1997)}]{Friedrich1997}
Friedrich R, Peinke J (1997).
\newblock \enquote{Description of a Turbulent Cascade by a {Fokker-Planck}
  Equation.}
\newblock \emph{Physical Review Letters}, \textbf{78}, 863--866.
\newblock \doi{10.1103/PhysRevLett.78.863}.

\bibitem[{Friedrich \emph{et~al.}(2011)Friedrich, Peinke, Sahimi, and Reza
  Rahimi~Tabar}]{Friedrich2011}
Friedrich R, Peinke J, Sahimi M, Reza Rahimi~Tabar M (2011).
\newblock \enquote{Approaching Complexity by Stochastic Methods: From
  Biological Systems to Turbulence.}
\newblock \emph{Physics Reports}, \textbf{506}(5), 87--162.
\newblock ISSN 0370-1573.
\newblock \doi{10.1016/j.physrep.2011.05.003}.

\bibitem[{Gardiner(2004)}]{Gardiner2004}
Gardiner CW (2004).
\newblock \emph{Handbook of Stochastic Methods}.
\newblock Third edition. Springer Verlag, Berlin.

\bibitem[{Hadjihosseini \emph{et~al.}(2014)Hadjihosseini, Peinke, and
  Hoffmann}]{Hadjihosseini2014}
Hadjihosseini A, Peinke J, Hoffmann N (2014).
\newblock \enquote{Stochastic Analysis of Ocean Wave States With and Without
  Rogue Waves.}
\newblock \emph{New Journal of Physics}, \textbf{16}(5), 053037.
\newblock \doi{10.1088/1367-2630/16/5/053037}.

\bibitem[{H\"anggi and Thomas(1982)}]{Hanggi1982}
H\"anggi P, Thomas H (1982).
\newblock \enquote{Stochastic Processes: Time-Evolution, Symmetries and Linear
  Response.}
\newblock \emph{Physics Reports}, \textbf{88}, 207--319.
\newblock \doi{10.1016/0370-1573(82)90045-X}.

\bibitem[{Honisch and Friedrich(2011)}]{Honisch2011}
Honisch C, Friedrich R (2011).
\newblock \enquote{Estimation of Kramers-Moyal Coefficients at Low Sampling
  Rates.}
\newblock \emph{Physical Review E}, \textbf{83}(6), 066701.
\newblock \doi{10.1103/PhysRevE.83.066701}.

\bibitem[{Kleinhans(2012)}]{Kleinhans2012}
Kleinhans D (2012).
\newblock \enquote{Estimation of Drift and Diffusion Functions from Time Series
  Data: A Maximum Likelihood Framework.}
\newblock \emph{Physical Review E}, \textbf{85}(2), 026705.
\newblock ISSN 1550-2376.
\newblock \doi{10.1103/physreve.85.026705}.

\bibitem[{Kleinhans \emph{et~al.}(2007)Kleinhans, Friedrich, W\"achter, and
  Peinke}]{Kleinhans2007d}
Kleinhans D, Friedrich R, W\"achter M, Peinke J (2007).
\newblock \enquote{Markov Properties in Presence of Measurement Noise.}
\newblock \emph{Physical Review E}, \textbf{76}(4), 041109.
\newblock \doi{10.1103/PhysRevE.76.041109}.

\bibitem[{Lamouroux and Lehnertz(2009)}]{Lamouroux2009}
Lamouroux D, Lehnertz K (2009).
\newblock \enquote{Kerner-based Regression of Drift and Diffusion Coefficients
  of Stochastic Processes.}
\newblock \emph{Physics Letters A}, \textbf{373}, 3507--3512.
\newblock \doi{10.1016/j.physleta.2009.07.073}.

\bibitem[{Lehle(2011)}]{Lehle2011}
Lehle B (2011).
\newblock \enquote{Analysis of Stochastic Time Series in the Presence of Strong
  Measurement Noise.}
\newblock \emph{Physical Review E}, \textbf{83}(2), 021113.
\newblock ISSN 1550-2376.
\newblock \doi{10.1103/physreve.83.021113}.

\bibitem[{Lehle(2013)}]{Lehle2013}
Lehle B (2013).
\newblock \enquote{Stochastic Time Series with Strong, Correlated Measurement
  Noise: {Markov} Analysis in $N$ Dimensions.}
\newblock \emph{Journal of Statistical Physics}, \textbf{152}(6), 1145--1169.
\newblock ISSN 1572-9613.
\newblock \doi{10.1007/s10955-013-0803-z}.

\bibitem[{Lemon(2006)}]{Lemon2006}
Lemon J (2006).
\newblock \enquote{\pkg{Plotrix}: A Package in the Red Light District of
  \proglang{R}.}
\newblock \emph{R-News}, \textbf{6}(4), 8--12.
\newblock
  \urlprefix\url{https://cran.r-project.org/doc/Rnews/Rnews_2006-4.pdf}.

\bibitem[{Lind \emph{et~al.}(2010)Lind, Haase, B{\"{o}}ttcher, Peinke,
  Kleinhans, and Friedrich}]{lind2010}
Lind PG, Haase M, B{\"{o}}ttcher F, Peinke J, Kleinhans D, Friedrich R (2010).
\newblock \enquote{Extracting Strong Measurement Noise from Stochastic Time
  Series: Applications to Empirical Data.}
\newblock \emph{Physical Review E}, \textbf{81}, 041125.
\newblock \doi{10.1103/PhysRevE.81.041125}.

\bibitem[{Lind \emph{et~al.}(2014)Lind, Herr\'aez, W\"achter, and
  Peinke}]{Lind2014}
Lind PG, Herr\'aez I, W\"achter M, Peinke J (2014).
\newblock \enquote{Fatigue Loads Estimation Through a Simple Stochastic Model.}
\newblock \emph{Energies}, \textbf{7}, 8279--8293.
\newblock \doi{10.3390/en7128279}.

\bibitem[{{R Core Team}(2015)}]{RCT2015}
{R Core Team} (2015).
\newblock \emph{\proglang{R}: A Language and Environment for Statistical
  Computing}.
\newblock R Foundation for Statistical Computing, Vienna, Austria.
\newblock \urlprefix\url{http://www.R-project.org/}.

\bibitem[{Renner \emph{et~al.}(2001)Renner, Peinke, and Friedrich}]{Renner2001}
Renner C, Peinke J, Friedrich R (2001).
\newblock \enquote{Experimental Indications for {Markov} Properties of
  Small-scale Turbulence.}
\newblock \emph{Journal of Fluid Mechanics}, \textbf{433}, 383--409.
\newblock \doi{10.1017/S0022112001003597}.

\bibitem[{Riera and Anteneodo(2010)}]{Riera2010}
Riera R, Anteneodo C (2010).
\newblock \enquote{Validation of Drift and Diffusion Coefficients from
  Experimental Data.}
\newblock \emph{Journal of Statistical Mechanics: Theory and Experiment},
  \textbf{2010}(04), P04020.
\newblock \doi{10.1088/1742-5468/2010/04/P04020}.

\bibitem[{Rinn \emph{et~al.}(2012)Rinn, Hei{\ss}elmann, W{\"{a}}chter, and
  Peinke}]{Rinn2012}
Rinn P, Hei{\ss}elmann H, W{\"{a}}chter M, Peinke J (2012).
\newblock \enquote{Stochastic Method for In-situ Damage Analysis.}
\newblock \emph{The European Physical Journal B}, \textbf{86}, 1--5.
\newblock ISSN 1434-6028.
\newblock \doi{10.1140/epjb/e2012-30472-8}.

\bibitem[{Risken(1996)}]{Risken1996}
Risken H (1996).
\newblock \emph{The Fokker-Planck Equation}.
\newblock Springer-Verlag.

\bibitem[{Sanderson(2010)}]{Sanderson2010}
Sanderson C (2010).
\newblock \enquote{Armadillo: An Open Source \proglang{C++} Linear Algebra
  Library for Fast Prototyping and Computationally Intensive Experiments.}
\newblock \emph{Technical report}, NICTA.
\newblock \urlprefix\url{http://arma.sourceforge.net/armadillo_nicta_2010.pdf}.

\bibitem[{Scholz \emph{et~al.}(2015)Scholz, Raischel, Lopes, Lehle, W\"achter,
  Peinke, and Lind}]{Scholz2015}
Scholz T, Raischel F, Lopes V, Lehle B, W\"achter M, Peinke J, Lind P (2015).
\newblock \enquote{Parameter-free Resolution of the Superposition of Stochastic
  Signals.}
\newblock \emph{submitted}.
\newblock \urlprefix\url{http://arxiv.org/abs/1510.07285}.

\bibitem[{Siegert \emph{et~al.}(1998)Siegert, Friedrich, and
  Peinke}]{Siegert1998}
Siegert S, Friedrich R, Peinke J (1998).
\newblock \enquote{Analysis of Data Sets of Stochastic Systems.}
\newblock \emph{Physics Letters A}, \textbf{243}, 275--280.
\newblock \doi{10.1016/S0375-9601(98)00283-7}.

\bibitem[{Stresing and Peinke(2010)}]{Stresing2010}
Stresing R, Peinke J (2010).
\newblock \enquote{Towards a Stochastic Multi-point Description of Turbulence.}
\newblock \emph{New Journal of Physics}, \textbf{12}(10), 103046.
\newblock \doi{10.1088/1367-2630/12/10/103046}.

\bibitem[{Van~Kampen(2007)}]{Kampen2007}
Van~Kampen N (2007).
\newblock \emph{Stochastic Processes in Physics and Chemistry}.
\newblock North-Holland Personal Library, third edition. Elsevier, Amsterdam.

\bibitem[{W\"achter \emph{et~al.}(2011)W\"achter, Milan, M\"ucke, and
  Peinke}]{Waechter2011}
W\"achter M, Milan P, M\"ucke T, Peinke J (2011).
\newblock \enquote{Power Performance of Wind Energy Converters Characterized as
  Stochstic Process: Applications of the Langevin Power Curve.}
\newblock \emph{Wind Energy}, \textbf{14}, 711--717.
\newblock \doi{10.1002/we.453}.

\bibitem[{Wilcoxon(1945)}]{Wilcoxon1945}
Wilcoxon F (1945).
\newblock \enquote{Individual Comparisons by Ranking Methods.}
\newblock \emph{Biometrics}, \textbf{1}, 80--83.
\newblock \doi{10.2307/3001968}.

\end{thebibliography}

\end{document}